\documentclass[conference]{IEEEtran}
\IEEEoverridecommandlockouts
\usepackage{cite}
\usepackage{amsmath,amssymb,amsfonts}
\usepackage{algorithmic}
\usepackage{graphicx}
\usepackage{textcomp}
\usepackage{xcolor}
\usepackage{balance}
\usepackage{hyperref}
\usepackage{subcaption}
\usepackage{multirow}
\usepackage{caption}
\def\BibTeX{{\rm B\kern-.05em{\sc i\kern-.025em b}\kern-.08em
    T\kern-.1667em\lower.7ex\hbox{E}\kern-.125emX}}
\begin{document}

\title{Evaluating the Practicality of\\Learned Image Compression}
\author{\IEEEauthorblockN{Hongjiu Yu$^1$, Qiancheng Sun\thanks{$^{\star}$ This work is done when the marked authors are interns at SenseTime.}$^{1\star}$, Jin Hu$^{1\star}$, Xingyuan Xue$^1$, Jixiang Luo$^1$, Dailan He$^1$, Yilong Li$^{1\star}$, Pengbo Wang$^{1\star}$,\\
Yuanyuan Wang$^1$, Yaxu Dai$^1$, Yan Wang\thanks{$^{\dagger}$ Corresponding author.}$^{12\dagger}$, Hongwei Qin$^1$}
\IEEEauthorblockA{\textit{SenseTime Research$^1$, Tsinghua University$^2$} \\
\{yuhongjiu, sunqiancheng, hujin, xuexingyuan, luojixiang, hedailan, liyilong, wangpengbo\}@sensetime.com\\ \{wangyuanyuan, daiyaxu, wangyan1, qinhongwei\}@sensetime.com, wangyan@air.tsinghua.edu.cn
}
}

\maketitle

\begin{abstract}
Learned image compression has achieved extraordinary rate-distortion performance in PSNR and MS-SSIM compared to traditional methods. However, it suffers from intensive computation, which is intolerable for real-world applications and leads to its limited industrial application for now. In this paper, we introduce neural architecture search (NAS) to designing more efficient networks with lower latency, and leverage quantization to accelerate the inference process. Meanwhile, efforts in engineering like multi-threading and SIMD have been made to improve efficiency. Optimized using a hybrid loss of PSNR and MS-SSIM for better visual quality, we obtain much higher MS-SSIM than JPEG, JPEG XL and AVIF over all bit rates, and PSNR between that of JPEG XL and AVIF. Our software implementation of LIC achieves comparable or even faster inference speed compared to jpeg-turbo while being multiple times faster than JPEG XL and AVIF. Besides, our implementation of LIC reaches stunning throughput of 145 fps for encoding and 208 fps for decoding on a Tesla T4 GPU for 1080p images. On CPU, the latency of our implementation is comparable with JPEG XL.  
\end{abstract}


\section{Introduction}
Learned image compression (LIC) has outperformed traditional methods like JPEG~\cite{jpeg} and BPG~\cite{bpg} in terms of PSNR and MS-SSIM. Its outstanding rate-distortion performance mostly derives from jointly training all parameters instead of handcrafted and independent module design. Hyperprior and context frameworks~\cite{imagecnn5, imagecnn7, minnen2020channel, elic} dramatically improve compression gain across all ranges of bit rates. However, traditional methods have their own merits for industrial applications including smaller memory footprint, less computational complexity and hardware-friendly characteristics. LIC has an extremely high demand for hardware resources. Large model size and intensive computation hinder real-world deployment of LIC. Furthermore, correct decoding across platforms is not guaranteed due to the non-determinism of model inference and error can propagate catastrophically in entropy decoding~\cite{balle2018integer}. All these obstacles restrict commercial deployment of LIC even though it obtains a better trade-off between bit rate and PSNR or MS-SSIM. In this paper, we report the performance of our optimized LIC implementation, which outperforms the fastest industrial-level codec implementation, i.e., jpeg-turbo~\cite{jpegturbo} regarding rate-distortion performance and visual quality while achieving comparable speed. Compared with the most competitive next-generation image codec JPEG XL~\cite{alakuijala2019jpeg}, our model shows superior compression performance with several times of advantage regarding inference speed.

To obtain a practical LIC model, we apply neural architecture search (NAS) and quantization to make LIC lighter and faster. NAS has been proved useful in finding efficient models for high-level vision tasks, such as object detection~\cite{chen2019detnas},  recognition~\cite{ding2020autospeech} and segmentation~\cite{cubuk2019autoaugment}, while researches on NAS for low-level tasks such as denoising, super-resolution and image compression are still in early stages. In this paper, we aim to design a more efficient network and hardware-friendly architecture for LIC to enable its industrial deployment. Hardware-aware NAS is dedicated to designing promising networks for various devices.

Besides, quantization has exhibited promising results for decreasing model size as well as accelerating inference. Methods for neural network quantization can be divided into two categories of quantization-aware training (QAT) and post-training quantization (PTQ). We exploit quantization not only to accelerate the inference of LIC but also to ensure the correctness of cross-platform decoding~\cite{balle2018integer}. 

The paper is organized as follows. Section \uppercase\expandafter{\romannumeral2} introduces related works. Section \uppercase\expandafter{\romannumeral3} gives an overview of our architecture and pipeline. Section \uppercase\expandafter{\romannumeral4} describes the details of our method. Experimental results and discussion are given in section \uppercase\expandafter{\romannumeral5}, and conclusion is drawn in section \uppercase\expandafter{\romannumeral6}.

\section{Related Work}
\subsection{NAS}
Search space, optimization methods and search strategy are the major components of NAS. In this paper, we follow the framework of BigNAS~\cite{yu2020bignas}, which only trains one big model and searches efficient sub-networks for direct deployment without additional training. For optimization methods, Yu \emph{et al.}~\cite{yu2018slimmable, yu2019universally} propose a slimmable neural network with parameter sharing which can adaptively prune output channels with little performance degradation. HyperNet~\cite{ha2016hypernetworks} trains an auxiliary network to generate weight parameters for the trainable model. For search strategy,  an effective evolution method~\cite{real2019regularized}  is proposed to obtain architecture with higher accuracy, and HyperBand~\cite{li2017hyperband} can assess candidate neural architectures to avoid redundant training which results in worse performance.  

\begin{figure}[tb]
    \centering
    \includegraphics[width=0.99\linewidth]{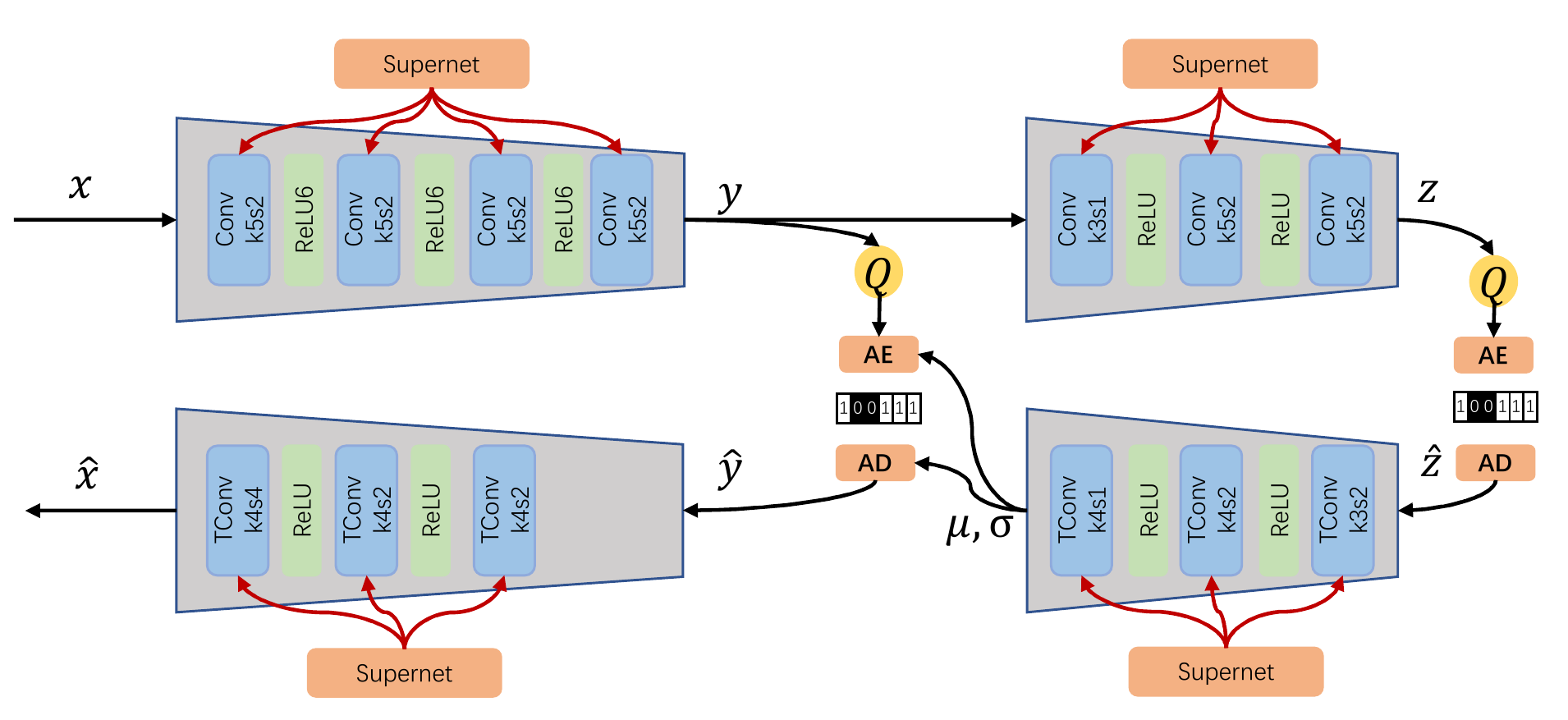}
    \caption{Framework of our architecture. $x, \hat{x}$ are the original image and the reconstructed one. $y, \hat{y}, z, \hat{z}$ are latent variables. $\mu, \sigma$ are mean and variance values of Gaussian distribution. $Q$ represents quantization. \textit{AE} and \textit{AD} are the arithmetic encoding and decoding. The blue blocks are (transposed) convolutional layers controlled by \textit{supernet} in Fig.~\ref{fig:supernet}, and $k5s2$ means $5 \times 5$ kernel with stride $2$. \textit{ReLU, ReLU6} are the activation functions.}
    \label{fig:framework}
\end{figure}

\subsection{Quantization}

% QAT and cross-platform...
Quantization-Aware Training (QAT)~\cite{choi2018pact, gong2019differentiable, esser2019learned} and Post-Training Quantization (PTQ)~\cite{nagel2020up, li2021brecq, wei2022qdrop} are two major categories of model quantization. QAT has attracted much attention because it compensates for the quantization error in an end-to-end optimization manner and achieves relatively better performance even in extremely low bit-width scenarios. Besides speeding up the model inference, the quantization techniques are also used in LIC for another purpose: eliminating the cross-platform inconsistency. Ball{\'e} \emph{et al.}~\cite{balle2018integer} first introduces a QAT-like integer-only arithmetic technique to determinize the hyperprior model. Sun \emph{et al.}~\cite{sun2021learned} instead adopts PTQ-based approaches and He \emph{et al.}\cite{he2022post} extends the deterministic inference to more elaborate joint autoregressive models.

\section{Architecture}
\begin{figure}[tb]
    \centering
    \includegraphics[width=0.95\linewidth]{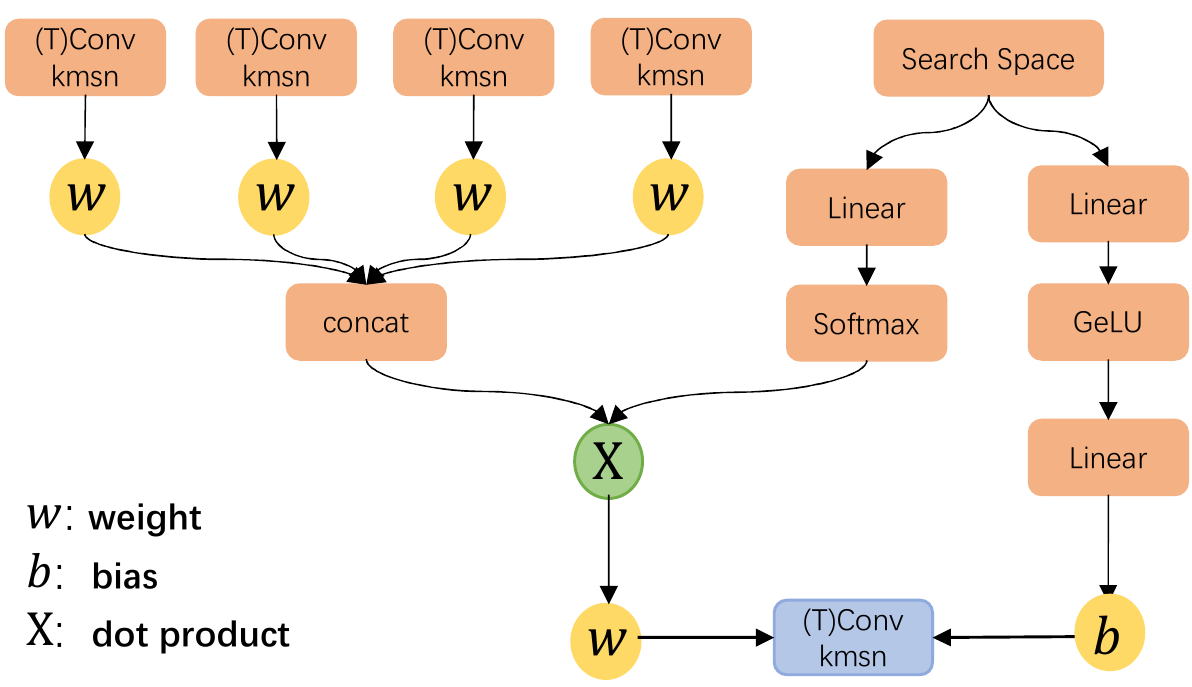}
    \caption{Supernet of our NAS architecture. $w, b$ are the weight and bias parameters. The notation $kmsn$ means $m \times m$ kernel with stride $n$. \textit{Search Space} is the combination of parameters for channels. \textit{Linear} is the linear convolutional layer. \textit{GeLU, Softmax} are the activation functions where $GeLU(x)=0.5 x(1+\tanh{[\sqrt{\frac{\pi}{2}} (x+0.004715 x^3)]})$, and $Softmax(x_i)=\frac{e^{x_i}}{\sum^l_{i=0} e^{x_i}}$, $l$ is the length of input vector $\mathrm{x}$. The blue block is the convolution adopted in Fig.~\ref{fig:framework}. }
    \label{fig:supernet}
\end{figure}

Our framework is shown in Fig.~\ref{fig:framework}. We follow the pipeline in ~\cite{imagecnn5} but use less complex activation functions and supernet to search the parameters of convolution layers, whose weight and bias are generated by supernet in Fig.~\ref{fig:supernet}. 

% \subsection{Supernet}

\section{Methodology}

\subsection{Conditional and Hardware-aware NAS}
Firstly we define the search space. The minimal and maximal numbers of parameters to be searched are given empirically, and the search stride is predefined for each convolutional layer. BigNAS can train several models at the same time, and the sandwich rule with combinations of ${min, random, random, max}$ is embedded to guarantee the effectiveness of sub-models, where ${min, random}$ and ${max}$ denote the sub-models with minimal, random and maximal number of parameters. After training, we can directly deploy sub-models for validation. 

For each convolutional layer in Fig.~\ref{fig:framework}, weight and bias parameters are generated by supernet in Fig.~\ref{fig:supernet}. A group of weights from the initial transposed convolution or convolution layers will be combined by \textit{concat} operation. Then the dot product of concatenated tensor and latent variables extracted from search space parameters by multiple layer perception (MLP), which consists of a single \textit{Linear} layer and a \textit{Softmax} layer, will be calculated to generate weight. Meanwhile, bias will be directly generated by latent variables from MLP, which consists of two \textit{Linear} layers and a \textit{GeLU} activation layer. Then the original parameters are replaced with generated weight and bias for training and testing. 

During the searching process, we balance the rate-distortion performance and computational complexity by taking FLOPS and inference latency into consideration. FLOPS is calculated to restrict the model size and a latency lookup table obtained from the statistical characteristics of various operations on specific hardware is used to constrain the inference latency. 

\subsection{Other Optimization}\label{otheropt}
Efforts have been made in engineering, resulting in a highly optimized software implementation. We leverage multi-threading and instruction-level parallelism to accelerate our implementation, reaching throughput higher than the reciprocal of latency. For entropy coding, multi-threading and AVX-512 are also used for acceleration.

\section{Experiments}
\subsection{Setup}\label{setup}
We train our model with 8000 images picked from ImageNet~\cite{deng2009imagenet} dataset and randomly crop each image to $256\times256$. For evaluation, we use 24 Kodak~\cite{1993kodak} images.

For training settings, we set learning rate to $1e-4$ and train the network for 2000 epochs. We follow~\cite{minnen2020channel} and use the mixed quantization trick to train our network. Channel numbers in main and hyper networks are obtained with NAS. In terms of quantization experiments, we apply LSQ~\cite{esser2019learned} to quantize all layers in our network following the general quantization rules of NVIDIA's TensorRT Framework. We use the full precision model to initialize weights. Then we train the network for 400 epochs with learning rate set to $1e-4$ and another 100 epochs with learning rate decreased to its $\frac{1}{10}$. 

For traditional codecs, we measure the latency for executing the whole pipeline including disk i/o and png format conversion for a fair comparison.
For JPEG XL~\cite{alakuijala2019jpeg}, we tune the distance parameter for rate-distortion trade-off and leave other parameters as default. For AVIF, we test with AOM codec by adjusting its quantization step tuned by PSNR unless otherwise specified. 

To test the average latency of our LIC model and traditional codecs, we use 100 1080p images captured from a video and measure its average processing time. For a fair comparison, when comparing latency between LIC model and jpeg-turbo~\cite{jpegturbo}, we test these images saved in \emph{bmp} file format while we use images saved in \emph{png} format when comparing LIC model, JPEG XL and AVIF. Unless otherwise mentioned, 
we test on a Intel(R) Core(TM) i9-10900X CPU @ 3.70GHz and a Tesla T4 GPU for 1080p dataset with png format.   
\subsection{Rate Distortion Performance}
\begin{figure}[tb]
\centering
    \begin{subfigure}[b]{0.45\textwidth}
        \centering
        \includegraphics[width=\linewidth]{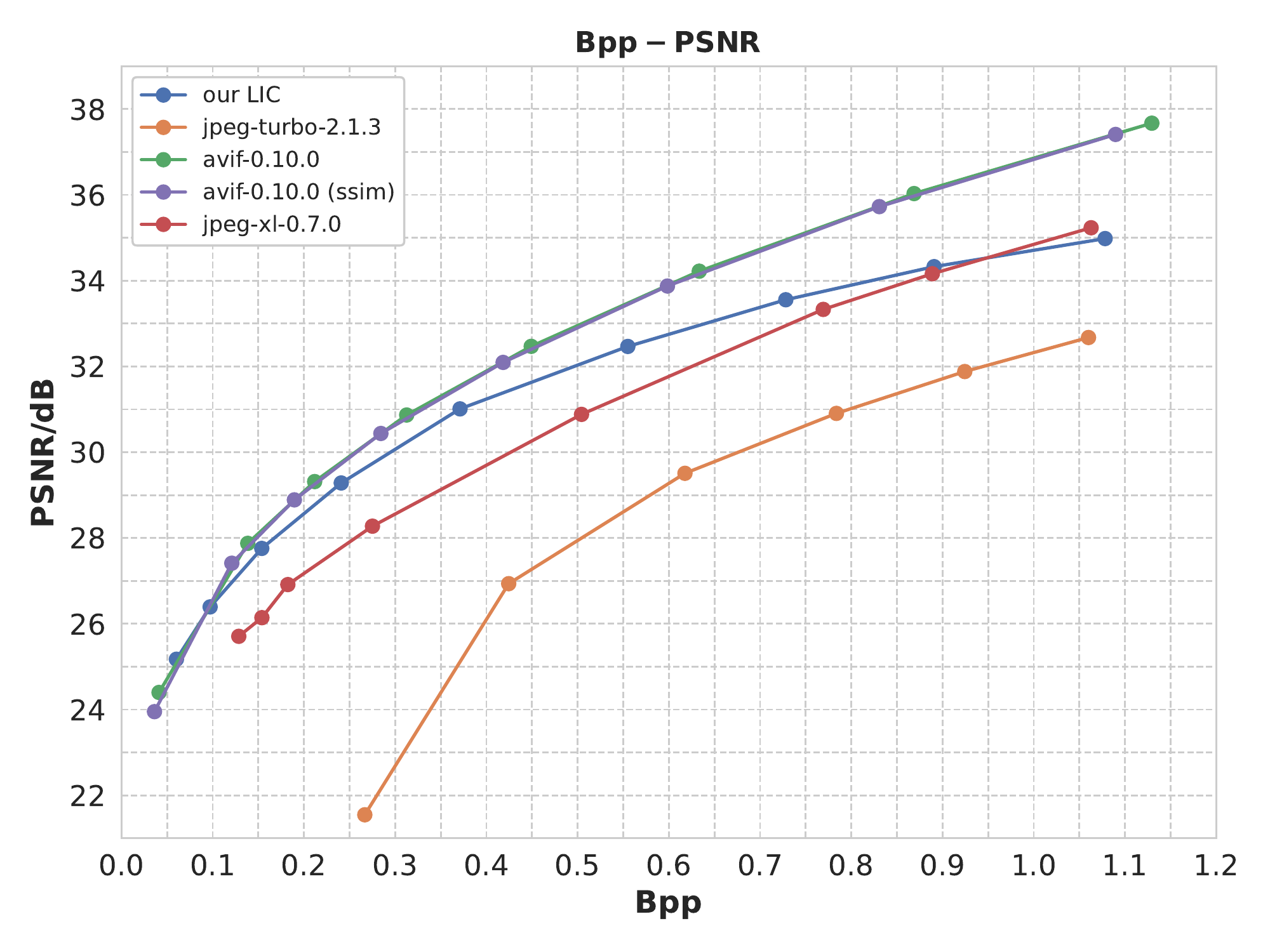}%
        \caption{BPP-PSNR curves of codecs}
        \label{fig:bpp-psnr}
    \end{subfigure}
    \\
    \begin{subfigure}[b]{0.45\textwidth}
        \centering
        \includegraphics[width=\linewidth]{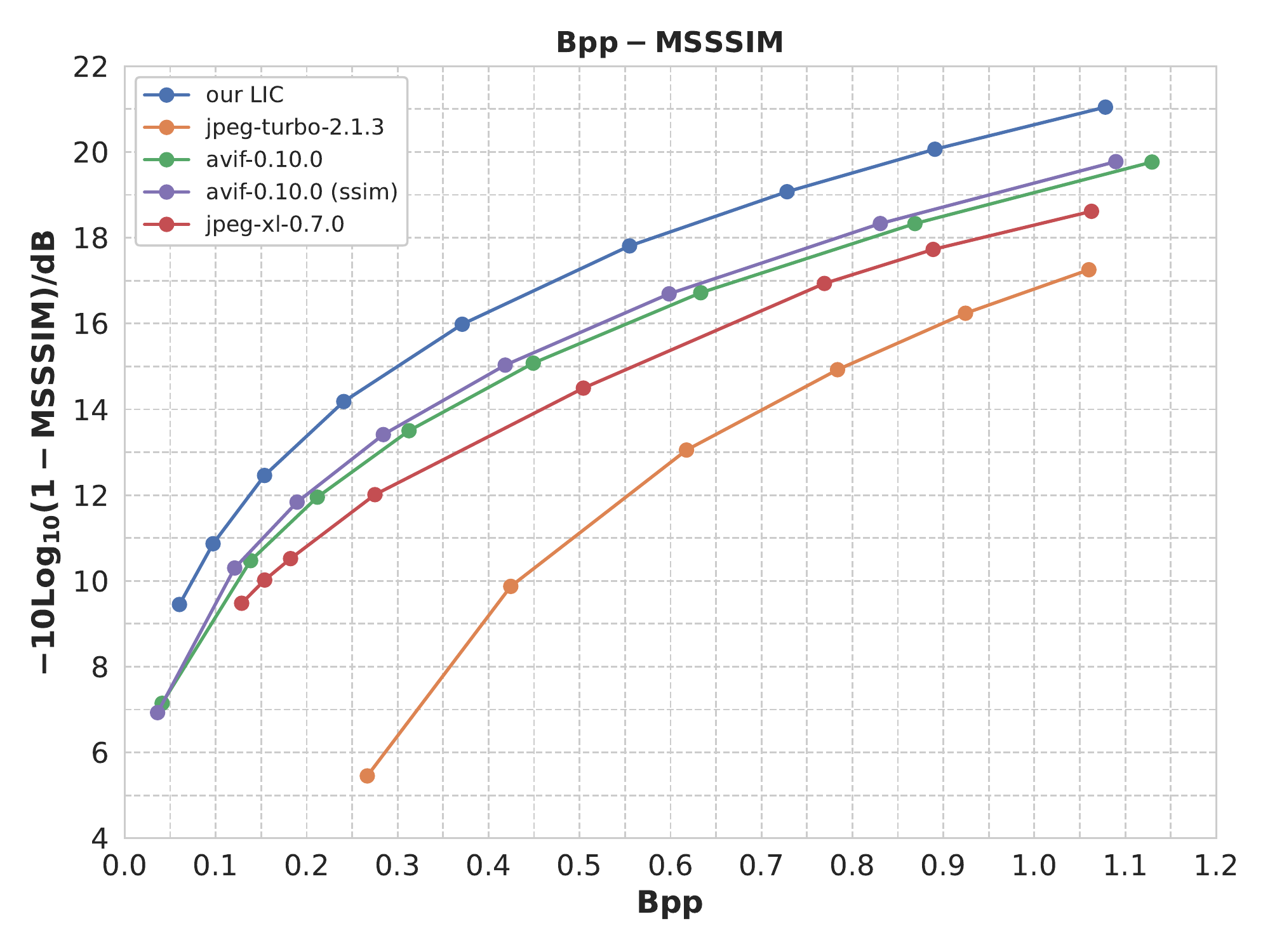}%
        \caption{BPP-MS-SSIM curves of codecs}
        \label{fig:bpp-msssim}
    \end{subfigure}
    \caption{Performance of our LIC model compared with traditional codecs with respect to bpp v.s. psnr and bpp v.s. ms-ssim. ``avif-0.10.0 (ssim)" denotes AVIF tuned by ssim.}
\end{figure}
As shown in Fig.~\ref{fig:bpp-psnr}, our LIC model achieves superior performance with a large margin regarding BPP-PSNR compared with jpeg-turbo, especially in low bit rates. It also defeats JPEG XL at low bit rates. Its performance is still comparable to JPEG XL when bpp exceeds 1.0. For AVIF, our LIC model is comparable at low bit rates but lags as bit rate increases.

Our model surpasses traditional codecs in terms of BPP-MS-SSIM, as shown in Fig.~\ref{fig:bpp-msssim}. MS-SSIM has a closer relationship with perceptual quality~\cite{cheng2019perceptual} and for deep networks, they tend to obtain a larger advantage over traditional codecs in terms of BPP-MS-SSIM, which can be verified from the fact that both LIC (Learned Image Compression) and DVC (Deep Video Compression) surpass traditional codecs in terms of BPP-MS-SSIM before BPP-PSNR~\cite{imagecnn7, lu2019dvc}. We follow the same rationale by using a hybrid loss of PSNR and MS-SSIM~\cite{sun2021hlic} to optimize perceptual quality, which leads to our model performing better than traditional codecs including AVIF tuned by ssim by a large margin in terms of BPP-MS-SSIM but worse than AVIF in terms of BPP-PSNR.
\subsection{Speed Analysis}
\begin{figure}[tb]
\centering
    \begin{subfigure}[b]{0.4\textwidth}
        \centering
        \captionsetup{width=80mm}
        \includegraphics[width=\linewidth]{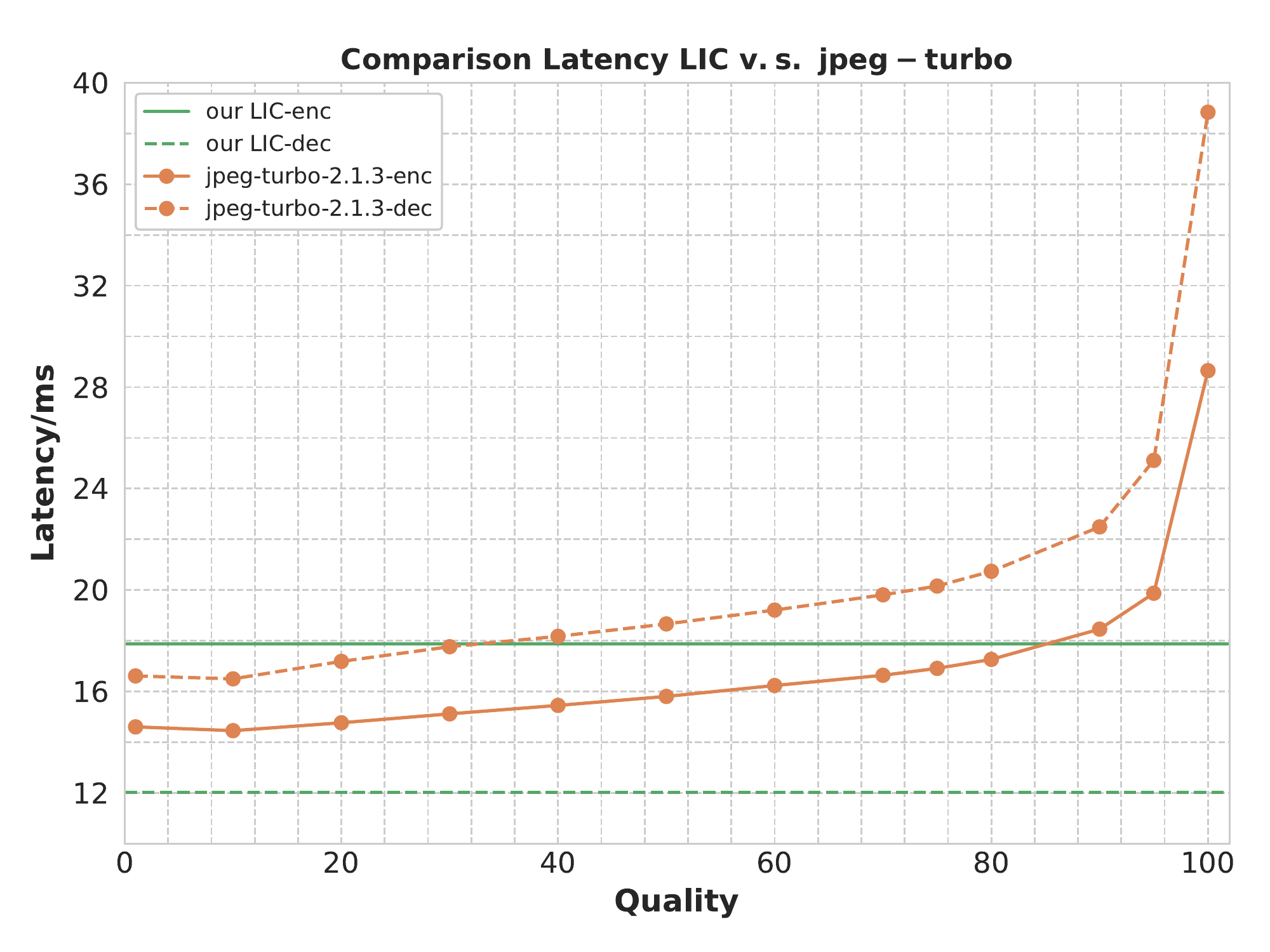}%
        \caption{Comparison of latency between our LIC model and jpeg-turbo for 1080p \emph{bmp} format images.}
        \label{fig:latency-turbo}
    \end{subfigure}
    \\
    \begin{subfigure}[b]{0.4\textwidth}
        \centering
        \captionsetup{width=80mm}
        \includegraphics[width=\linewidth]{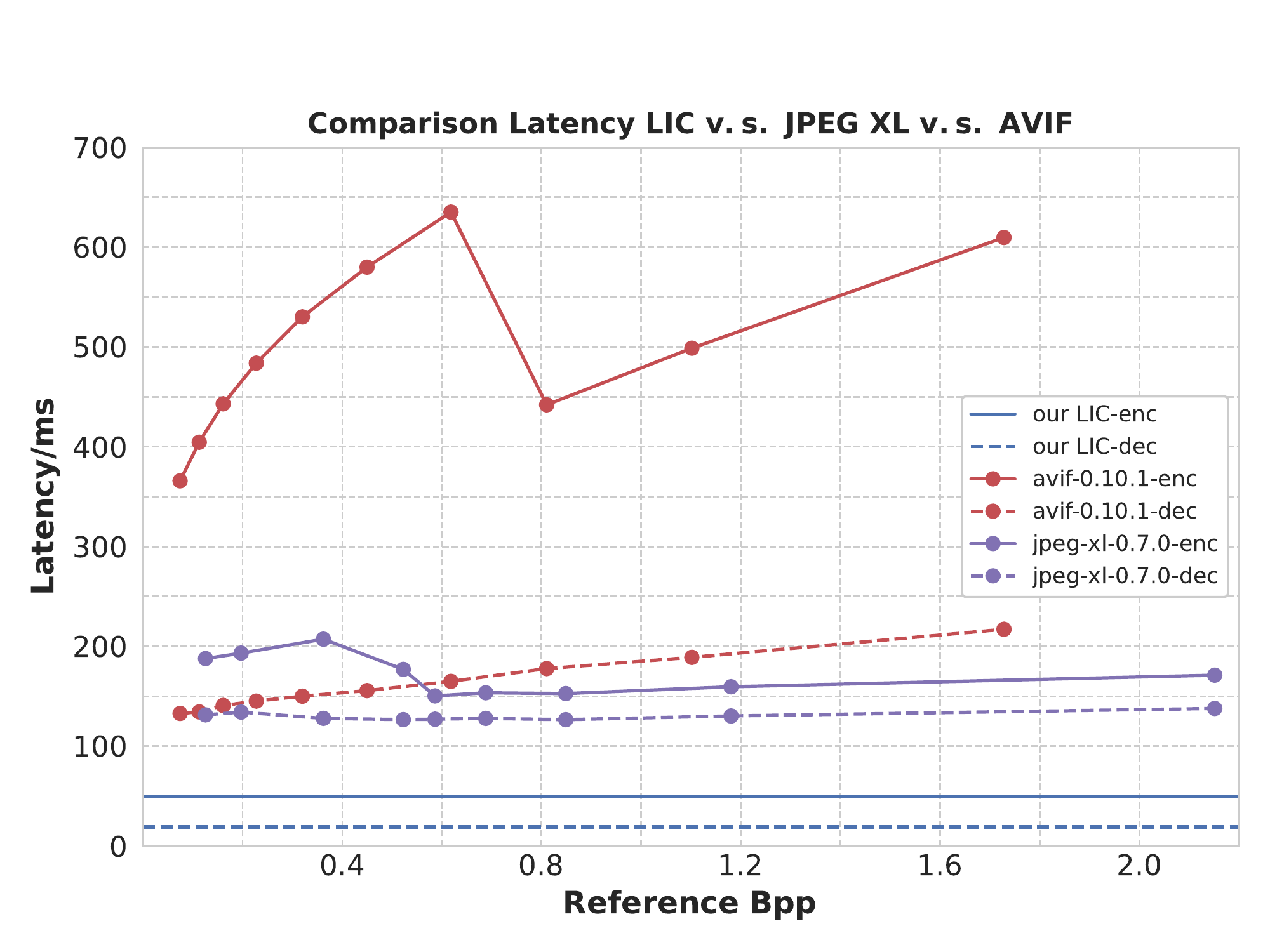}%
        \caption{Comparison of latency between our LIC model, AVIF and JPEG XL for 1080p \emph{png} format images.}
        \label{fig:latency-xl-avif}
    \end{subfigure}
\caption{Latency comparison of our LIC model and traditional codecs. 
``Quality" stands for jpeg qp while ``Reference Bpp" stands for bit rates on 1080p dataset for JPEG XL and AVIF.
}
\label{fig:latency}
\end{figure}

We compare the inference latency of our LIC model with traditional codecs and the results are shown in Fig.~\ref{fig:latency}. Our model achieves \textbf{18/12ms} for encoding/decoding latency, which is at most 20\% higher in encoding latency and at least 35\% lower in decoding latency than jpeg-turbo with \emph{bmp} format input. Compared with JPEG XL and AVIF, our LIC model is several times faster, reaching \textbf{50/19ms} for encoding and decoding processes respectively. To further demonstrate the practicality of our implementation of the LIC model, we also test its latency on a less expensive GPU (NVIDIA GeForce GTX 1660 SUPER) and a less powerful CPU (Intel(R) Core(TM) i7-8700 CPU @ 3.20GHz). Encoding/decoding latency is \textbf{58/25ms}. In addition, we have also tested the latency of our LIC model on CPU, reaching encoding/decoding latency of \textbf{199/284ms, comparable to JPEG XL} as shown in Fig.\ref{fig:latency-xl-avif}. We leave the optimization of CPU deployment as future work.

In terms of throughput, with techniques mentioned in Section \ref{otheropt}, when processing images in parallel, our implementation reaches stunning \textbf{145/208fps} for \emph{bmp} format input and \textbf{126/199fps} when input is saved in \emph{png} format for encoding and decoding processes. On a less expensive GPU (NVIDIA GeForce GTX 1660 SUPER), our model still reaches throughput of \textbf{92/82fps}.

We also report latency of each subnetwork in our model in Table.~\ref{table:latency} which is obtained using TensorRT's trtexec benchmarking tool with an input of resolution $1088\times1920$.
\begin{table}
\begin{center}
\begin{tabular}{ c |c c c c}
\hline \hline
 \textbf{GPU Type}& $g_a$ & $g_s$ &$h_a$&$h_s$ \\
 \hline
 \textbf{Tesla T4} & 7.25&	5.48&	1.15&	1.17 \\  
 \textbf{GeForce GTX 1660 SUPER} & 10.08 & 10.75 & 1.68 & 2.80   \\
 \hline
\end{tabular}
\caption{Latency of each subnetwork in our model. $g_a$, $g_s$, $h_a$ and $h_s$ above refer to analysis and synthesis transforms of main and hyperprior autoencoders.}
\label{table:latency}
\end{center}
\end{table}

\subsection{Memory Analysis}
\begin{figure}[tb]
\centering
    \begin{subfigure}[t]{0.23\textwidth}
        \centering
        \includegraphics[width=\linewidth]{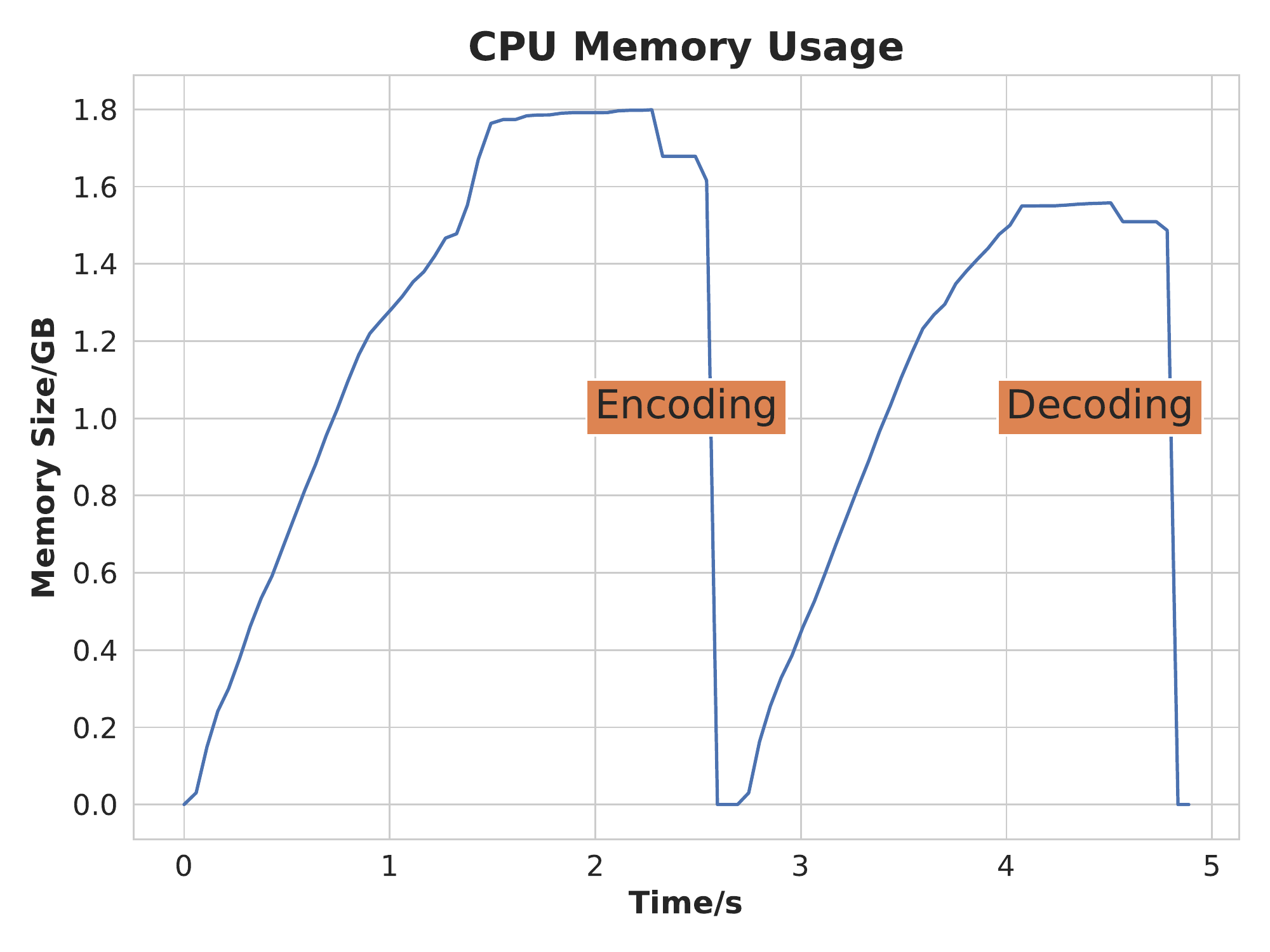}%
        \label{fig:cpu_mem}
    \end{subfigure}
    \begin{subfigure}[t]{0.23\textwidth}
        \centering
        \includegraphics[width=\linewidth]{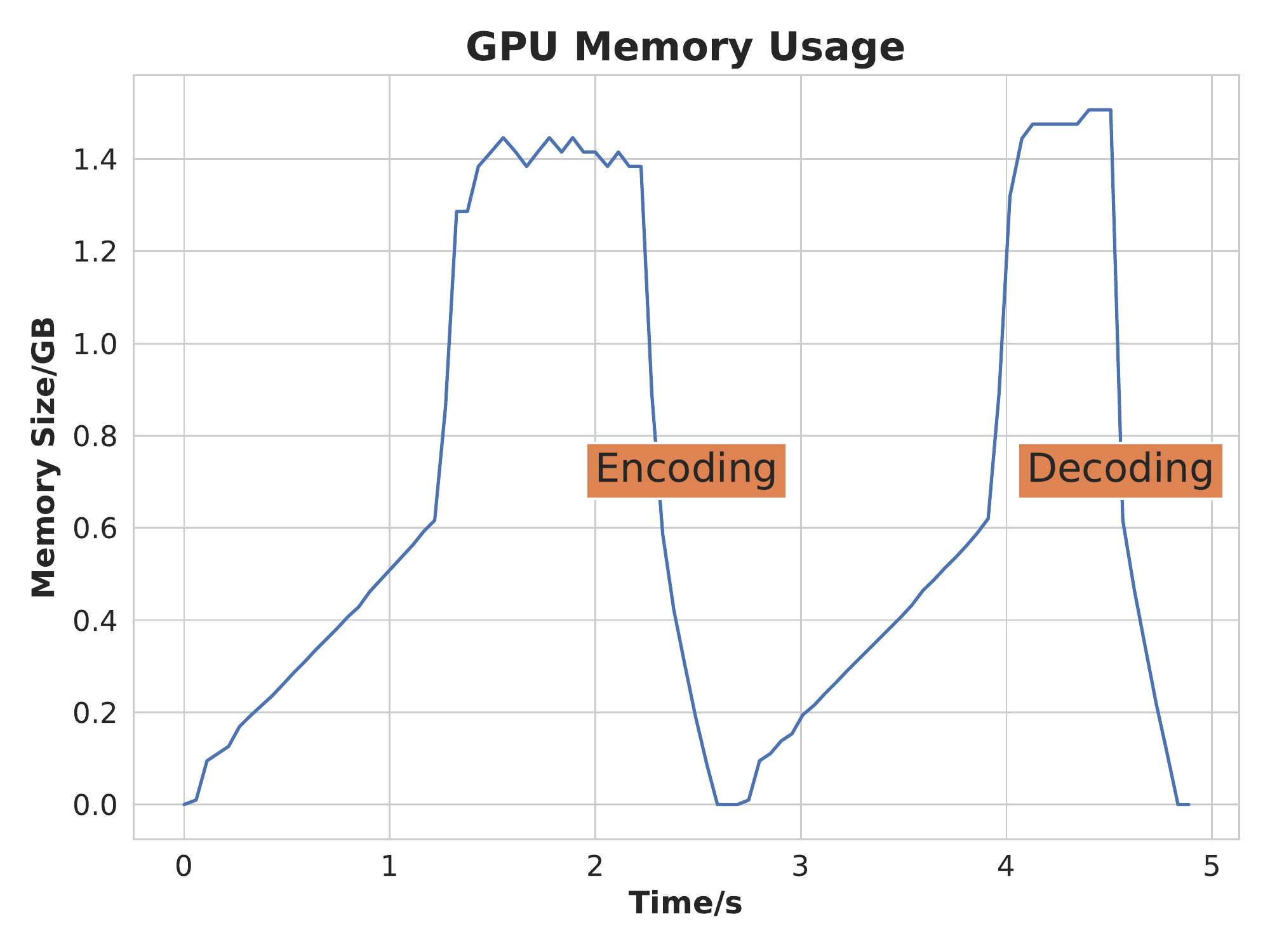}%
        \label{fig:gpu_mem}
    \end{subfigure}
    \caption{CPU and GPU memory usage during encoding and decoding.}
    \label{fig:mem}
\end{figure}

We show the memory footprint of our model during encoding and decoding processes of 100 1080p images in Fig.~\ref{fig:mem}, peaking at 1.8GB and 1.5GB for CPU and GPU respectively. 

\subsection{Visual Quality}
\begin{figure}[tb]
\centering
    \begin{subfigure}[t]{0.15\textwidth}
        \centering
        \includegraphics[width=\linewidth]{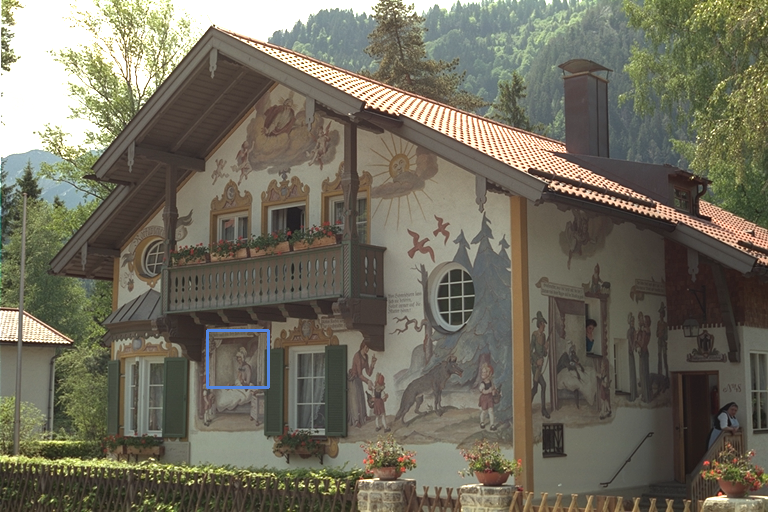}%
        \caption{Original image}
    \end{subfigure}
    \\
    \begin{subfigure}[t]{0.11\textwidth}
        \centering
        \includegraphics[width=\linewidth]{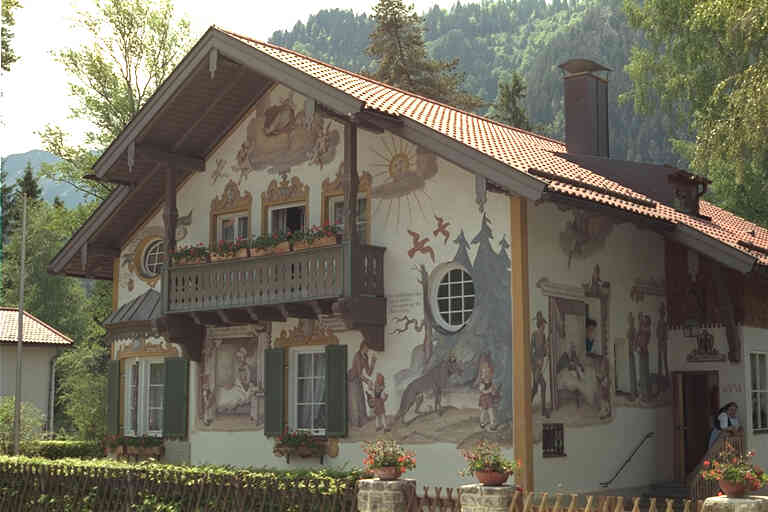}%
        \caption{JPEG \\44.4KB}
    \end{subfigure}
    \begin{subfigure}[t]{0.11\textwidth}
        \centering
        \includegraphics[width=\linewidth]{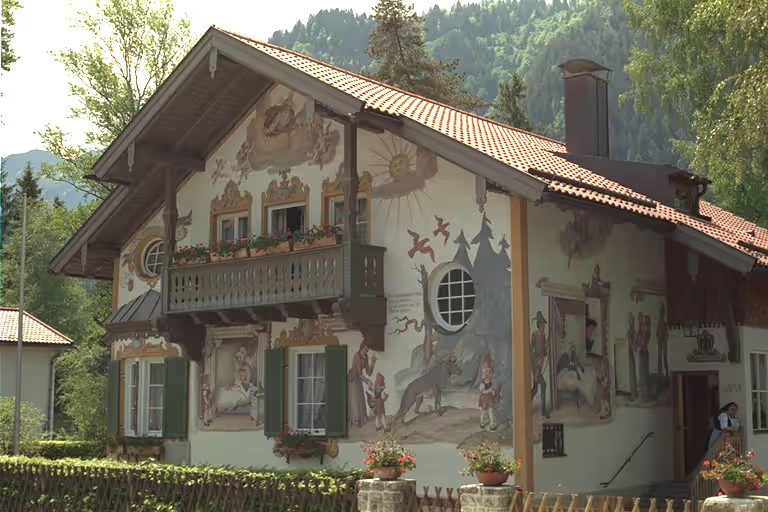}%
        \caption{AVIF \\43.75KB}
    \end{subfigure}
    \begin{subfigure}[t]{0.11\textwidth}
        \centering
        \includegraphics[width=\linewidth]{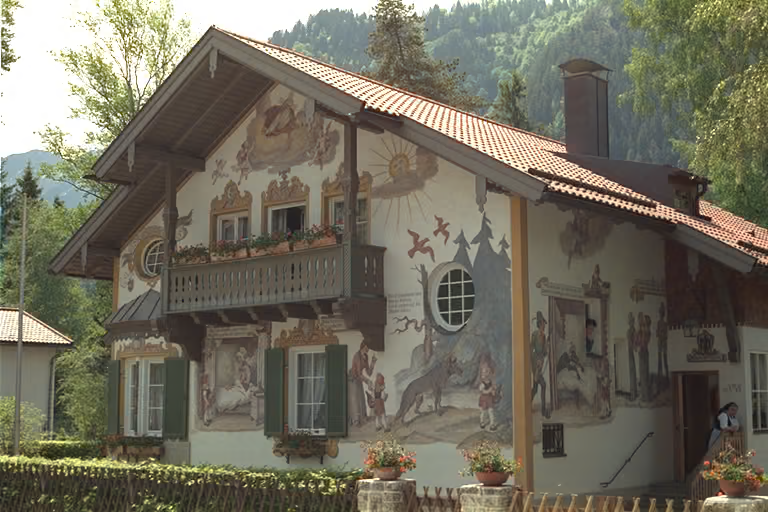}%
        \caption{JPEG XL \\44.35KB}
    \end{subfigure}
    \begin{subfigure}[t]{0.11\textwidth}
        \centering
        \includegraphics[width=\linewidth]{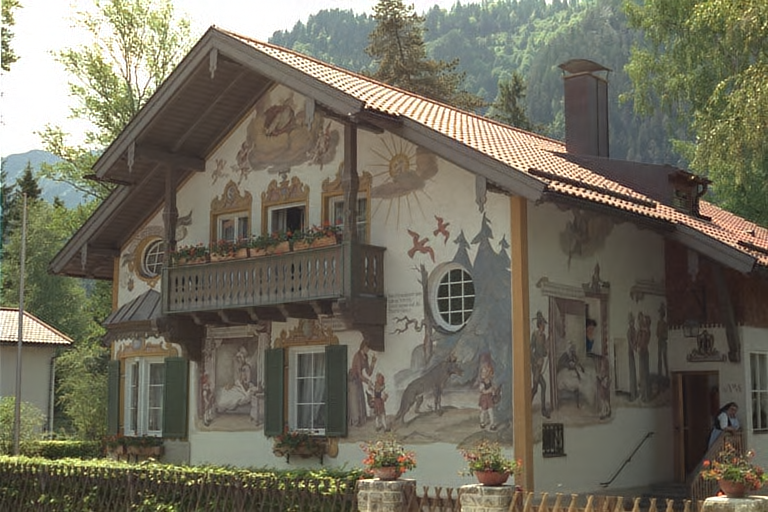}%
        \caption{our LIC \\44.44KB}
    \end{subfigure}
    \\
    \begin{subfigure}[t]{0.11\textwidth}
        \centering
        \includegraphics[width=\linewidth]{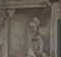}%
    \end{subfigure}
    \begin{subfigure}[t]{0.11\textwidth}
        \centering
        \includegraphics[width=\linewidth]{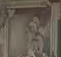}%
    \end{subfigure}
    \begin{subfigure}[t]{0.11\textwidth}
        \centering
        \includegraphics[width=\linewidth]{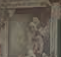}%
    \end{subfigure}
    \begin{subfigure}[t]{0.11\textwidth}
        \centering
        \includegraphics[width=\linewidth]{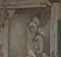}%
    \end{subfigure}
    \caption{Visual quality comparison of kodim24.png~\cite{1993kodak}. Images on the third row are obtained by zooming in the blue square box region of images above.}
    \label{fig:kodim24}
\end{figure}

In this subsection, we compare the visual quality of our LIC model with that of traditional codecs on Kodak~\cite{1993kodak} dataset. As can be seen in Fig.~\ref{fig:kodim24}, the image compressed by JPEG~\cite{jpeg} displays blocking artifacts while the image compressed by AVIF is oversmoothed, losing details. Image compressed by JPEG XL~\cite{alakuijala2019jpeg} suffers from blocking artifacts, similar to JPEG, but to a lesser extent. The image compressed by our LIC model, on the other hand, shows better perceptual quality.

\subsection{NAS}
 For NAS, we reduce about $10\%$ inference time while improving rate-distortion performance. %Then we present the original and searched number of output channel for $g_a, h_a, h_s, g_s$. 
 We get lower bpp ($0.7281 \rightarrow 0.7278$), with higher PSNR ($33.5550 \rightarrow 34.0997$) and higher MS-SSIM ($0.9876 \rightarrow 0.9880$), but with lower latency ($15.2059 \rightarrow 13.9574$) as shown in Table.~\ref{tab:Bignas}.

\begin{table}[]
\setlength{\tabcolsep}{3pt}
\begin{tabular}{ccccc}
\hline \hline
Structure   & $g_a$          & $h_a$       & $h_s$       & $g_s$     \\ \hline
Origin      &48,96,112,176& 176,246,176 & 246,176,176 & 112,96,3  \\ 
NAS      & 32,120,104,220 & 248,224,256 & 236,200,220 & 112,112,3 \\ \hline \hline
Performance & BPP            & PSNR        & MS-SSIM     & Lantency(ms)  \\ \hline
Origin      & 0.7281         & 33.5550     & 0.9876      & 15.2059   \\ 
NAS      & 0.7278         & 34.0997     & 0.9880      & 13.9547   \\ \hline
\end{tabular}
\caption{Output channels of each (de)convolution block in $g_a$, $g_s$, $h_a$ and $h_s$  and rate-distortion performance with latency are listed above.}\label{tab:Bignas}
\end{table}

\subsection{Quantization}
\begin{table}[]
\centering
\setlength{\tabcolsep}{6pt}
\begin{tabular}{c|ccc}
\hline \hline
model   & bpp         & psnr     &ms-ssim       \\ \hline
FP32      &0.728 & 33.55 & 0.9876 \\ 
Quantized      & 0.704     & 32.91    & 0.9839 \\
$g_a$-$h_a$-Quantized      &0.716	&33.49&	0.9873 \\\hline

\end{tabular}
\caption{BPP-PSNR-MS-SSIM of quantized models. ``Quantized" stands for the model with all layers quantized, evaluated on a NVIDIA GeForce GTX 1660 SUPER GPU. ``$g_a$-$h_a$-Quantized" denotes the model with its main and hyper encoders quantized, evaluated in pytorch's fake quantize mode.} 
\label{table:quant}
\end{table}
We show the rate-distortion performance of the quantized models in Table.~\ref{table:quant}. We notice that quantizing main and hyper encoders barely hurt performance while quantizing all layers leads to a minor loss in rate-distortion performance. As for speedup on a real device, quantization brings about approximately 20\% to 35\% increase in throughput (images processed in parallel) on a NVIDIA GeForce GTX 1660 SUPER GPU, from 92/82 fps to 114/112 fps.

\section{Conclusion}
We evaluate the practicality of learned image compression by optimizing and assessing rate-distortion performance, latency, throughput, memory footprint and visual quality. We compare these results with those of traditional codecs. LIC has comparable or even faster inference speed compared with the fastest industrial codec, i.e., jpeg-turbo, while being superior in terms of rate-distortion performance and perceptual quality. Besides, LIC achieves astonishing throughput of 145/208 fps for 1080p images, making it applicable for industrial scenarios. In addition, NAS and quantization show promising results for obtaining more efficient and more effective LIC models. All these results show that learned image compression is becoming ready for industrial applications. Future works will be devoted to optimizing LIC for mobile devices.

\balance
{\small
\bibliographystyle{IEEEtran}
\bibliography{ms}
}

\end{document}